\documentclass[a4paper]{cas-sc}

\usepackage[authoryear,longnamesfirst]{natbib}
\usepackage{graphicx}
\usepackage{amsmath}
\usepackage{amssymb}
\usepackage{booktabs}
\usepackage{algorithm}
\usepackage{algorithmic}
\usepackage{amssymb}
\usepackage{multirow}
\usepackage{booktabs}
\usepackage{diagbox}
\usepackage{color,xcolor}
\usepackage{tabularray}
\usepackage{wrapfig}
\usepackage[capitalize]{cleveref}
\crefname{section}{Sec.}{Secs.}
\Crefname{section}{Section}{Sections}
\Crefname{table}{Table}{Tables}
\crefname{table}{Tab.}{Tabs.}
\def\tsc#1{\csdef{#1}{\textsc{\lowercase{#1}}\xspace}}
\tsc{WGM}
\tsc{QE}
\usepackage{xcolor}

\begin{document}
\let\WriteBookmarks\relax
\def\floatpagepagefraction{1}
\def\textpagefraction{.001}

\shorttitle{Impart}    

\shortauthors{Zhao et al}  

\title [mode = title]{Impart: An Imperceptible and Effective Label-Specific Backdoor Attack}  



%

\author[1]{Jingke Zhao}



\ead[url]{hi_simon@tju.edu.cn}

\credit{ Conceptualization, Methodology, Software, Writing - Original Draft}

\affiliation[1]{organization={Tianjin University},
            addressline={92 Weijin Road}, 
            city={Tianjin},
            postcode={300072}, 
            country={China}}

\author[1]{Zan Wang}



\ead[url]{wangzan@tju.edu.cn}

\credit{Supervision, Writing - Review & Editing}

\affiliation[2]{organization={Nanyang Technological University},
            addressline={50 Nanyang Avenue}, 
            postcode={639798}, 
            country={Singapore}}

\cortext[1]{Corresponding author}


\author[2]{Yongwei Wang}
\ead[url]{yongweiw@ece.ubc.ca}
\credit{Writing - Review & Editing}
\author[1]{Lanjun Wang}[
    orcid=0000-0002-7696-5330
]

\ead[url]{wang.lanjun@outlook.com}
\credit{Project administration, Supervision, Writing - Review & Editing}


\cormark[1]

\begin{abstract}
   Backdoor attacks have been shown to impose severe threats to real security-critical scenarios. Although previous works can achieve high attack success rates, they either require access to victim models which may significantly reduce their threats in practice, or perform visually noticeable in stealthiness. Besides, there is still room to improve the attack success rates in the scenario that different poisoned samples may have different target labels (a.k.a., the all-to-all setting).  In this study, we propose a novel imperceptible backdoor attack framework, named Impart, in the scenario where the attacker has no access to the victim model.
Specifically, in order to enhance the attack capability of the all-to-all setting, we first propose a label-specific attack. Different from previous works which try to find an imperceptible pattern and add it to the source image as the poisoned image, we then propose to generate perturbations that align with the target label in the image feature by a surrogate model. In this way, the generated poisoned images are attached with knowledge about the target class, which significantly enhances the attack capability.   
   We conduct experiments on three benchmark datasets and five widely used defense mechanisms. 
   Experiments show that Impart achieves successful attacks and high imperceptibility in five image quality metrics. 
   For example, our method outperforms existing works with an average attack success rate 13\% in the all-to-all setting on CIFAR-100 while keeping highly imperceptible with average visual quality improvements from 34.24dB to 40.45dB in PSNR. 
   Additionally, we demonstrate that Impart can successfully bypass existing effective defense methods.
\end{abstract}


\begin{highlights}
\item We propose a novel backdoor attack framework, Impart, where the attacker uses a surrogate model to generate effective backdoor examples in the scenario where the attacker does not have access to the model information. 
\item we first propose a label-specific attack, where the generated backdoor examples are associated with the target label before the backdoor attack which significantly enhances the attack capability of the backdoor attack.
\item Our method Impart outperforms existing works with an average attack success rate 13\% in the all-to-all setting on CIFAR-100 while keeping highly imperceptible with average visual quality improvements from 34.24dB to 40.45dB in PSNR. 
\end{highlights}


\maketitle










\section{Introduction}
\label{sec:intro}
Deep Neural Networks (DNNs) have achieved remarkable success in the past few years and they have been adopted in different applications (e.g., image classification~\citep{he2016deep}, speech recognition~\citep{xiong2016achieving}, game playing and natural language processing~\citep{silver2016mastering, devlin2018bert}). However, with the deepening research on several real security-critical scenarios, recent works show that even the state-of-the-art deep learning methods are vulnerable to backdoor attacks~\citep{gu2017badnets, barni2019new, cheng2021deep, li2021invisible, CHENG2023109629}. In backdoor attacks, an attacker injects a trigger into the victim model in the training process. The victim model performs normally as a benign model in the inference phase when the inputs are benign images. However, once the victim model is fed an input image with the backdoor trigger, the victim model behaves as the attacker predetermined. {In the backdoor attack, there are two typical types of attack settings~\citep{li2022backdoor}: one is to poison different target labels (a.k.a., \textit{all-to-all}), and the other is to poison one target label (a.k.a., \textit{all-to-one}). }

Recent research on the backdoor attack for deep learning has focused on generating poisoned images that lead to misclassification results while keeping imperceptibility. LIRA~\citep{doan2021lira} and WB~\citep{doan2021backdoor} have achieved effective and imperceptible backdoor attacks. However, they assume that the attacker has full access to the model information (e.g., model architecture, and model parameters), which significantly reduces their threats in practice. Meanwhile, for existing black-box backdoor methods that do not utilize the information about the victim model~\citep{cheng2021deep, li2021invisible, nguyen2021wanet, wang2022bppattack}, the generated poisoned images are inevitably visually noticeable. 

Moreover, compared with the all-to-one setting, the all-to-all setting has not been well studied. We observe that there is still an unsatisfied attack success rate in the all-to-all setting.
{For example, WaNet~\citep{nguyen2021wanet}, an advanced black-box backdoor attack method, achieves a 99.56\% high ASR in the all-to-one setting on the CIFAR-10 dataset, yet it only has a 94\% ASR in the all-to-all setting. Even typical white-box methods, such as LIRA~\citep{doan2021lira} and WB~\citep{doan2021backdoor}, can only achieve about 94\% ASR in the all-to-all setting}. 
In addition, the all-to-all setting is more valuable to be investigated. First, the attacker can attack multiple classes simultaneously by injecting a type of backdoor trigger, which is much more useful and flexible in practice. Second, the all-to-all setting attacks are more difficult to be detected for their complicated target shifting, and therefore, are more serious compared with the all-to-one attacks~\citep{li2022backdoor}. Besides, as described in~\citet{doan2022marksman}, repeatedly injecting different trigger patterns for all the target classes is not feasible because it will lead to a much larger model perturbation and significantly degrade the clean data accuracy. There were only a few studies specifically designed on the all-to-all\citep{li2022backdoor} setting, yet how to better design the all-to-all attack remains underexplored.
It is still a crucial challenge to achieve an \textit{imperceptible and effective backdoor attack} for the all-to-all setting in the scenario where the attacker does not have access to the model information.  
First, the human visual system has different perception sensitivities to different regions~\citep{eckert1998perceptual, wang2022reaching} and different colors~\citep{zhao2020towards, wang2021perception}. For example, distortions around the edges are easier to be noticed than those in the texture regions. Another example is that even when identical perturbations are added in all three channels (RGB), the perturbations are more easily noticed in the green channel~\citep{li2021invisible}. However, some of the previous studies fail to consider this human visual sensitivity ~\citep{wang2022bppattack, nguyen2021wanet, li2021invisible}. For example, \citet{nguyen2021wanet} considers distorting the whole image, but not the texture regions. Besides, most existing studies try to manipulate the source image by some transformations involving randomness~\citep{nguyen2020input, li2021invisible, nguyen2021wanet, wang2022bppattack}. That is to say, the triggers are label-agnostic. Therefore, in order to achieve successful poison, there is a need for large amounts of transformations that damages imperceptibility as demonstrated in \cref{experiments}. In addition, backdoor attacks aim to make samples with those triggers output as the designed target label. However, due to the randomness, the victim model is hard to learn a {mapping} from triggers to the target label. 
To tackle these challenges, in this study, different from previous works that try to obtain an imperceptible pattern and add it to the source image as the poisoned image, we propose to generate perturbation that aligns with the target label in the image feature by a surrogate model. Specifically, in order to promote the learning of the model, we propose a {label-specific attack} in which a generated poisoned image is associated with the target label. Namely, we utilize the information of the target label to generate a backdoor trigger for a specific image. To summarize, we propose a new attack framework named \textbf{Impart}. Firstly, we train a surrogate model before the backdoor attack. The surrogate model is used to fit the image feature. Then, the surrogate model combined with the target label and learned image feature to generate triggers. In this way, the backdoor images (i.e., embed the generated triggers into the source image) are associated with the target label in the image feature before implementing the backdoor attack. Finally, we use backdoor images to train the victim model which enhances the mapping between the backdoor image and the target label. 

Our main contributions can be summarized as:
\begin{itemize}
    \item We propose a novel backdoor attack framework, Impart, where the attacker uses a surrogate model to generate effective backdoor examples in the scenario where the attacker does not have access to the model information. 
    \item To our best knowledge, we first propose a label-specific attack, where the generated backdoor examples are associated with the target label before the backdoor attack which significantly enhances the attack capability of the backdoor attack.
    \item We empirically evaluate Impart on three benchmark datasets and demonstrate that the proposed method achieves {high} imperceptibility, meanwhile outperforming existing methods in the all-to-all setting. For example, Impart outperforms existing works with an average attack success rate of 13\% in the all-to-all setting on CIFAR-100 while keeping highly imperceptible with average visual quality improvements from 34.24dB to 40.45dB in PSNR. 
    \item We further evaluate Impart under five widely used defense mechanisms and demonstrate Impart can successfully bypass all of them.
\end{itemize}
\begin{figure*}[htbp]

\centering
\includegraphics[width=13cm]{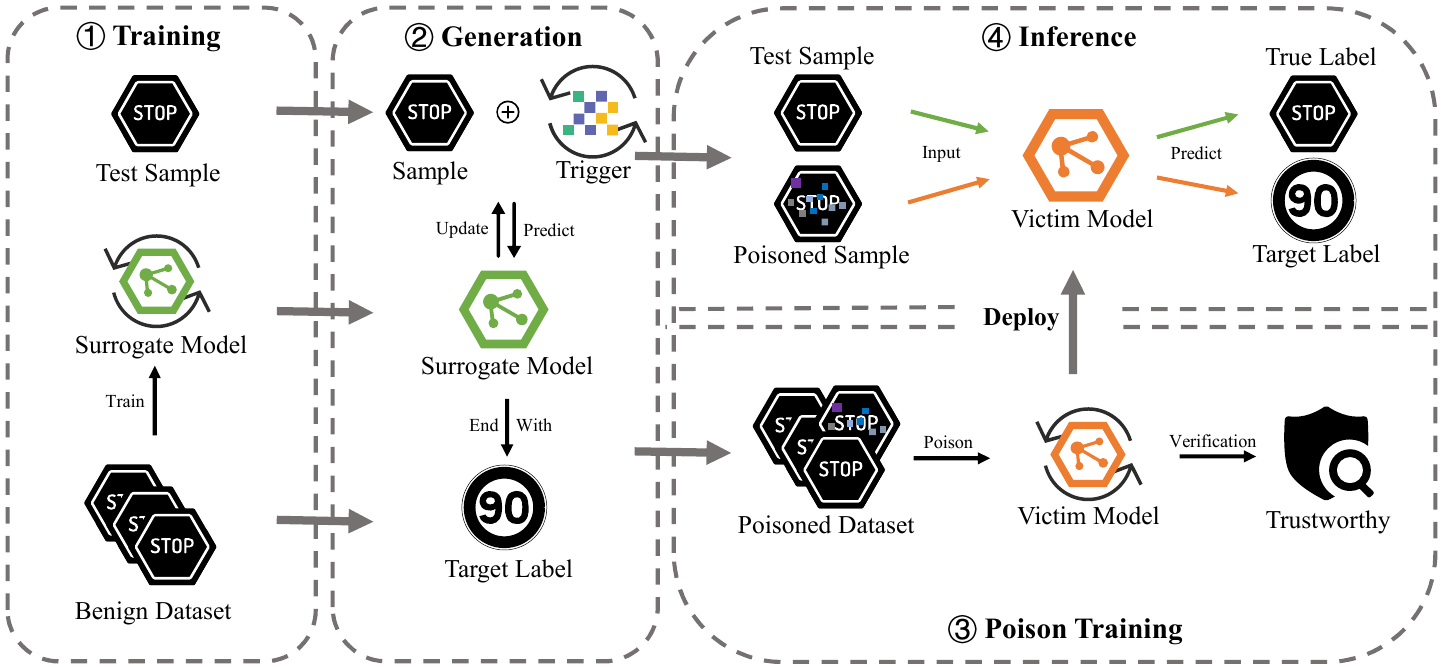}
\caption{{The framework of the proposed Impart method consists of four phases.} In phase \textcircled{1}, we train a surrogate model. In phase \textcircled{2}, we generate poisoned training data and poisoned test data using the feature fitter. In phases \textcircled{3} and \textcircled{4}, we poison and test the victim model respectively.}
\label{framework}
\end{figure*}
\section{Related Work}


As our method is motivated by adversarial attacks, we first investigate adversarial attack methods in \cref{sec:adversarial attack}, where we mainly focus on imperceptible adversarial attacks. Then, some {state-of-the-art} backdoor attack approaches are analyzed in \cref{sec:backdoor attack}.
\subsection{Adversarial Attacks}
\label{sec:adversarial attack}
Adversarial attacks arise as a severe threat in the model inference phase ~\citep{moosavi2016deepfool,yuan2019adversarial,pei2017deepxplore, MA2023109363}, where the attacker attempts to induce misclassification of a model by manipulating the inputs. For different purposes of attacks, the attacker can perform a targeted or non-targeted attack. In the targeted attack setting, the adversary can flexibly attack any target label.
Typically, \citep{papernot2016limitations} proposes an adversarial attack method in the white-box setting to generate perturbations using gradient descent with the $\ell_2$ regularization in the targeted setting. PerC-AL~\citep{rony2019decoupling} achieves state-of-the-art performance in invisible attacks. PerC-AL replaces the original penalty $\ell_2$ with a perceptual color difference metric CIEDE2000~\citep{luo2001development}. Besides, the implementation of PerC-AL decouples the joint optimization by alternately updating the perturbations with respect to either classification loss or perceptual color difference. However, it required time-consuming large epochs to find a perturbation that is both effective and imperceptible because of the alternate optimization process. Additionally, while the adversarial examples generated by PerC-AL {are} imperceptible on average, there are still some special examples that have poor visual quality. 

\subsection{Backdoor Attacks}
\label{sec:backdoor attack}
This section provides a brief overview of existing backdoor attacks, which can be classified into three categories: patch-based trigger attacks, input-specific trigger attacks, and adversarial examples-based attacks.  Our approach is more related to the last two categories.

\noindent\textbf{Patch-based Trigger.}
BadNets~\citep{gu2017badnets} is a typical patch-based backdoor attack method, which uses a pattern of bright pixels lying in the bottom right corner of
the image. \citep{Liu2018TrojaningAO} generates a trojan trigger by inversing the neurons. Then, several works~\citep{barni2019new, chen2017targeted, liu2020reflection} started to explore an effective pattern while keeping some stealthiness. They suggested that the poisoned images should look as identical as possible to the source image under human inspection. Unfortunately, the patch is still clearly noticeable in existing works.

\noindent\textbf{Input-Specific Trigger Attacks.}
To achieve better stealthiness, \citep{nguyen2020input} implemented an input-aware trigger generator by proposing a diversity loss. ISSBA~\citep{li2021invisible} proposed to generate
sample-speciﬁc invisible triggers through an encoder-decoder network, which encodes a speciﬁed string into benign images. Until recently, WaNet~\citep{nguyen2021wanet} and BppAttack~\citep{wang2022bppattack} have achieved an unnoticeable modification to some degree by traditional image warping and compression. As mentioned in \cref{sec:intro}, all of the above works ignore how the model behaves and how the human visual system perceives, leading to the added perturbation large enough to be learned by the model and perceived by human eyes. Besides, in previous works, the poisoned data are randomly distributed with respect to the target label before the training process.

\noindent\textbf{Adversarial Examples-Based Attacks.}
AdvDoor~\citep{zhang2021advdoor} is most related to our method. AdvDoor abstracts a fixed universal adversarial perturbation by combining multiple input-specific perturbations generated by the adversarial attack. However, it breaks up the advantage (i.e., revealing the defect of models) of adversarial examples because of the fixed perturbation. Another issue of universal adversarial perturbation is that the imperceptibility is worse than the input-specific perturbation. Our proposed method Impart is different from it. Specifically, we generate input-specific perturbations and the generated perturbation are designed for the target label.

\noindent\textbf{Arbitrary Label Attacks.} Recently,  \citet{doan2022marksman} propose a new backdoor attack Marksman Backdoor (MB) just like the adversarial attack that the adversarial can flexibly attack any target label during inference phases. They proposed a class-condition trigger generation function, and alternately update the generation function and victim model while fixing the other one during backdoor training. However, it is still a white-box method that significantly reduces the threat in reality. Besides, the MB is only a special all-to-one attack that the adversarial can flexibly attack any target. Therefore, as described in \cref{sec:intro}, how to achieve an imperceptible and effective backdoor attack for the all-to-all setting in the scenario where the attacker does not have access to the model information is still a crucial challenge.

\section{Threat Model}

The threat model has two aspects: one is the attack scenario (\cref{sec:attack scenario}), and the other one is the specific attack objective (\cref{sec:attack objective}).
\subsection{Attack Scenario}
\label{sec:attack scenario}
In this study, we focus on the scenario where a victim user uploads her/his dataset, model, and training schedule to an untrusted third-party platform (e.g., Google Cloud) and train their model. The attacker (i.e., the malicious platform) modifies the dataset and training schedule during the training process. Then, the victim user gets the poisoned model. After evaluating the model with clean test data, the victim user deploys it into the production environment. Specifically, we assume that the attackers have access to the training set, and can modify the training schedule. All of these settings are following previous works~\citep{nguyen2021wanet,wang2022bppattack}. However, in order to make our attack more practical, we assume the attackers have no information about the victim model architecture and model parameters. 
\subsection{Attack Objective}
\label{sec:attack objective}
The main goal of our method is to produce a high attack success rate (ASR) after injecting a backdoor into the victim model while keeping a comparable accuracy for benign data. Specifically, the victim model which is trained in the poisoned dataset behaves normally in benign data as a benign model that is trained in the benign dataset. Once the attacker feeds inputs with the trigger into the victim model, the victim model behaves as the attacker pre-determined. Besides, imperceptibility is another goal that we need to achieve, which requires the poisoned data as identical as possible to benign data under human visual inspection.

\section{Method}

In this section, we first provide a formal problem formulation of the backdoor attack in \cref{sec:problem formulation}, and then present the detail of our method Impart in \cref{sec:impart}.
\subsection{Problem Formulation}
\label{sec:problem formulation}
As in the standard supervised images classification task, the goal is to learn the parameters $\theta$ of a mapping function $f_\theta:\mathcal{X} \rightarrow \mathcal{C}$ from the training dataset $D_{tr}=\{(x_i,y_i)\;|\; x_i \in \mathcal{X}, y_i \in \mathcal{C}, i=1,\dots, N\}$.

In the backdoor attack, the attacker aims to produce a poisoned subset
$D_p=\{(T(x_i), \eta(y_i))\;|\; (x_i, y_i)\in D_s \subset D_{tr}, \eta(y_i)\in \mathcal{C} \}$, 
where $D_s$ is a subset of $D_{tr}$ and $|D_s| = M=\rho N$, $\rho$ is the poison ratio, $T$ is a backdoor transformation function, and $\eta$ is a target label function. In the backdoor attack, there are two typical types of target label functions~\citep{li2022backdoor},
\begin{enumerate}
    \item \textit{all-to-all}: $\eta(y) = (y+1)\; mod\; |C|$, where the true label is one-shifted.
    \item \textit{all-to-one}: $\eta(y) = c$, where c is a fixed constant label;
\end{enumerate}
The poisoned model is trained on the dataset 
$D=D_p \cup D_r$, where $D_r=D_{tr}\setminus D_s$.
The poisoned model with new parameters $\theta^*$ is required to satisfy:
\begin{align}
     & f_{\theta^*}(x)=y \\
     & f_{\theta^*}(T(x))=\eta(y)
\end{align}
For the traditional perturbation-based method, the transformation functions $T$ can be defined as:
\begin{equation}
        T(x) = x + g(x), \|g(x) \| < \epsilon \,,
\end{equation}
where $g$ is an imperceptible perturbation generative function.

Differently, for the label-specific attack, we are trying to generate poisoned images associated with the target label. This can be formally described by:
\begin{equation}
        T(x,c) = x + t(x,c) \,,
\end{equation}
where $t(x,c)$ is a surrogate model which is used to fit the image feature and then combined with the target label and learned image feature to generate triggers. In this way, the image feature in the class $c$ can be embedded in the poisoned image $T(x,c)$ and enhance the backdoor attack capability. Besides, we expect the perturbation generative function $t(x, c)$ to align with the characteristic of the human visual system. This can be formally described by:
\begin{equation}
\begin{aligned}
    &f_\theta(T(x, c))= \eta(y)\\
   s.t. & \; d\big(T(x, c), x\big) \leq \epsilon\, ,
\end{aligned}
\end{equation}
where $d$ is a distance metric quantifying the human visual system, $\epsilon$ is the upper bound of perpetual color difference. We detail the solution in the next section.


\subsection{Impart}
\label{sec:impart}
\cref{framework} shows the framework of our method Impart {that consists of four phases}. {First,} we train a surrogate model using the identical dataset as the victim model \textcircled{1}. Then, by utilizing the target label and the knowledge of the surrogate model, we generate poisoned examples that are aligned to the target label in the image feature \textcircled{2}. Next, using the generated poisoned data to train the victim model \textcircled{3}. After the model is deployed into the production environment, the adversary applies phase \textcircled{2} to generate poisoned test data to attack \textcircled{4}. 

\noindent\textbf{Training a Surrogate Model.}
The first step of our method is to train the surrogate model using the identical clean dataset as the victim model. Formally, given a training data $D_{tr}$, we train the surrogate model $g_\vartheta$ by minimizing the following loss function:
\begin{equation}
        \mathop{\min}\limits_{\vartheta}  \sum_{i=1}^N \mathcal{L}(g_\vartheta(x_i),\; y_i)\,, 
\end{equation}
where $\mathcal{L}(\cdot)$ denotes a loss function and $g$ is the surrogate model.

\noindent\textbf{Generating Poisoned Data.}
After training the auxiliary surrogate model $g_\vartheta$, we will generate the poisoned data subset $D_p$. Our goal is to generate poisoned data that are as close to the source data as possible while achieving a high ASR. Besides, to improve the effectiveness of poisoning data, the generated perturbations need to relate to the target label. This can be solved by the following optimization problem:
\begin{equation}
\begin{aligned}
        \mathop{\min}\limits_{\delta}\; \mathcal{L}\big(g_\vartheta(x+\delta),\eta(y)\big)+ \gamma\cdot \|\delta\|_2        + \|\Delta  E_{00}(x+\delta,x)\|_2\,.
\end{aligned}
\label{loss}
\end{equation}
In \cref{loss}, we aim to generate imperceptible perturbation $\delta$ with the constraint of $\ell_2$ regularization and the perpetual color difference metric $\Delta E_{00}$. 

For the first item $\mathcal{L}\big(g_\vartheta(x+\delta),\eta(y)\big)$, where $\mathcal{L(\cdot)}$ is a standard cross-entropy (CE) loss function. It tries to find a $\delta$ which leads to the poisoned data $x+\delta$ aligned with the target label $\eta(y)$ in the image feature. 

Considering that the human visual system is sensitive different from color to color. Therefore, using the second item $\|\Delta  E_{00}(x+\delta,x)\|_2$ to align the feature of human visual system. We use the latest standard formula developed
by the International Commission on Illumination (CIE), which can be calculated as~\citep{luo2001development}: 
\begin{equation}
\begin{aligned}
    &\Delta E_{00}=\sqrt{
    \bigg(\frac{\Delta L'}{k_L S_L}\bigg)^2 +\bigg(\frac{\Delta C'}{k_C S_C}\bigg)^2+\bigg(\frac{\Delta H'}{k_H S_H}\bigg)^2
    } + \Delta R\\
    &\Delta R = R_T\cdot \frac{\Delta C'}{k_C S_C}\cdot \frac{\Delta H'}{k_H S_H}\,.
\end{aligned}
\label{ciede}
\end{equation}
\cref{ciede} calculates the color difference in Lab color space, where $\Delta L', \Delta C', \Delta H'$ denote the distance of two inputs image in channel Lightness, Chroma and Hue respective, and $\Delta R$ is an interactive item between channel chroma and hue. The $k_L,k_C,k_H$ are based on application scenarios, and $S_L, S_C, S_H$ act as compensation to better align with the human visual system. The parameter settings follow the paper~\citep{luo2001development}.

The third item $\gamma \cdot \|\delta\|_2$ is to further limit the update range of $\delta$ and speed up the optimization process. Specifically, using the item $\|\delta\|$, the update range of $\delta$ will be limited to small values. In this way, the imperceptibility of generated poisoned data are further improved. Besides, the number of iterations for finding an imperceptible perturbation are also greatly reduced. Additionally, the item $\|\gamma\|$ ensures that the perturbation of all the generated poisoned data are minor and uniform to some degree, which is a primary reason for being able to pass existing defense methods. We experimentally demonstrate this in \cref{gamma}.

Following the PerC-AL~\citep{zhao2020towards}, we solve the optimization problem by alternately updating the perturbations with respect to either classification loss and $\ell_2$ or perceptual color difference. Specifically, we first update the perturbation $\delta$ with classification loss and $\ell_2$ by gradient descent, then reduce the perpetual color difference of perturbations by gradient descent until the perturbation $\delta$ loses its effect. Then, move back and forth.

\noindent\textbf{Poisoning the Victim Model.}
Then, using the dataset $D=D_p\cup D_r$, where $D_r = D_{tr} \setminus D_s$ to poison the model $f_\theta$, which can be formulated to:
\begin{equation}
    \theta^* = \mathop{argmin}\limits_{\theta}  \sum_{i=1}^N \mathcal{L}\big( f_\theta(x_i), y_i\big)\,,
\end{equation}
where $(x_i,y_i) \in D$.

\section{Experiments}
\label{experiments}
In this section, we conduct attack experiments on three benchmark datasets and five image quality metrics to evaluate the attack's effectiveness and imperceptibility. Besides, we select five widely used defense mechanisms to evaluate whether our proposed method is resistant to them. Additionally, we investigate the influence of hyperparameters $\rho$, and $\gamma$.
\subsection{Experimental Setup}
\label{sec:exp setup}
\begin{table}[htbp]
\centering
\caption{\centering The details of datasets and network architectures.}
\label{dataset}
\begin{tabular}{rrrc} 

\toprule
Dataset    & \#Classes                  & VM  & ACC(\%)\\ 
\hline
CIFAR-10    & 10                  & PreResNet18   & 94.81   \\
GTSRB       & 43                  & PreResNet18   & 99.23   \\
CIFAR-100   & 100           & ResNet18       & 76.19   \\
\bottomrule
\end{tabular}

\end{table}
\noindent\textbf{Datasets and Models.}
We conduct our method in three commonly used datasets: CIFAR-10~\citep{krizhevsky2009learning}, GTSRB~\citep{stallkamp2012man} and CIFAR-100~\citep{krizhevsky2009learning}. Following the previous work~\citep{nguyen2021wanet}, we consider the blend networks with datasets. Specifically, as for the victim model(VM), we use Pre-activation ResNet18~\citep{he2016identity} for CIFAR-10 and GTSRB datasets, and ResNet18 for CIFAR-100 dataset. For the surrogate model(SM) used to fit the image feature, we select four existing models GoogleNet~\citep{szegedy2015going}, EfficientNetB0, EfficientNetB1~\citep{tan2019efficientnet} and ResNet34~\citep{he2016identity} to test the effect of our framework. More detailed information is in \cref{dataset}.

\noindent\textbf{Baselines.}
We compare our proposed method Impart with WaNet~\citep{nguyen2021wanet} and Bpp~\citep{wang2022bppattack} because both of them are state-of-the-art methods in the backdoor attack. Besides, they can be applied to the attack scenario which has no access to the victim model. 
We implement WaNet and Bpp on CIFAR-100 by using the default parameters stated in the original papers and reference their results of CIFAR-10 and GTSRB directly. For WaNet, the poisoned ratio is set at 10\% and the noise ratio is set at 20\%\footnote{The noise ratio in WaNet is used to control noise data generation for bypassing some defenses, and the negative ratio in Bpp plays the same role. However, Impart does not need extra data for defense purposes.}. For Bpp, the poisoned ratio and the negative ratio are both set at 20\%. 
{It is noted that for WaNet and Impart, the poisoned ratio is 10\%, but for Bpp, the poisoned ratio is 20\%\footnote{In experiments by the official code \url{https://github.com/RU-System-Software-and-Security/BppAttack}, we observed Bpp could not converge with a poisoned ratio as 10\%.}. 
Therefore, Bpp is easier to achieve a high attack success rate than the others since Bpp uses many more poisoned examples.
} 
For both of WaNet and Bpp, we train the network using  SGD optimizer with weight decay $5\times 10^{-4}$. The initial learning rate is set to 0.01 with a learning rate decay of factor 10 after every 100 epochs.

\begin{table*}[htbp]
\centering

\caption{Attack effectiveness in the all-to-all setting. For each dataset, four rows represent the SM as GoogleNet, EfficientNetB0, EfficientNetB1, and ResNet34, respectively. $*$ means the poisoned ratio is 20\%, and the others is 10\%.
}
\label{acc for all2all}

\begin{tblr}{
  row{odd} = {c},
  row{2} = {c},
  row{4} = {c},
  row{8} = {c},
  row{12} = {c},
  cell{1}{1} = {r=2}{},
  cell{1}{2} = {r=2}{},
  cell{1}{3} = {c=3}{},
  cell{1}{7} = {c=3}{},
  cell{3}{1} = {r=4}{},
  cell{3}{4} = {r=4}{},
  cell{3}{5} = {r=4}{},
  cell{3}{8} = {r=4}{},
  cell{3}{9} = {r=4}{},
  cell{6}{2} = {c},
  cell{6}{3} = {c},
  cell{6}{7} = {c},
  cell{7}{1} = {r=4}{},
  cell{7}{4} = {r=4}{},
  cell{7}{5} = {r=4}{},
  cell{7}{8} = {r=4}{},
  cell{7}{9} = {r=4}{},
  cell{10}{2} = {c},
  cell{10}{3} = {c},
  cell{10}{7} = {c},
  cell{11}{1} = {r=4}{},
  cell{11}{4} = {r=4}{},
  cell{11}{5} = {r=4}{},
  cell{11}{8} = {r=4}{},
  cell{11}{9} = {r=4}{},
  cell{14}{2} = {c},
  cell{14}{3} = {c},
  cell{14}{7} = {c},
  hline{1,15} = {-}{0.08em},
  hline{2} = {3-5,7-9}{},
  hline{3,7,11} = {-}{},
}
            & SM             & BA(\%)           &         &                  &  & ASR(\%)          &         &                \\
            &                & Impart           & WaNet   & $\text{Bpp}^*$   &  & Impart           & WaNet   & $\text{Bpp}^*$ \\
CIFAR-$10$  & GoogleNet      & $94.24$          & $94.43$ & $\mathbf{94.73}$ &  & $96.35$          & $93.36$ & $94.32$        \\
            & EfficientNetB0 & $94.15$          &         &                  &  & $\mathbf{97.56}$ &         &                \\
            & EfficientNetB1 & $94.21$          &         &                  &  & $97.40$          &         &                \\
            & ResNet34       & $94.43$          &         &                  &  & $97.04$          &         &                \\
GTSRB       & GoogleNet      & $98.03$          & $99.39$ & $\mathbf{99.46}$ &  & $98.90$          & $98.32$ & $99.29$        \\
            & EfficientNetB0 & $99.00$          &         &                  &  & $\mathbf{99.34}$ &         &                \\
            & EfficientNetB1 & $99.14$          &         &                  &  & $99.21$          &         &                \\
            & ResNet34       & $98.62$          &         &                  &  & $98.97$          &         &                \\
CIFAR-$100$ & GoogleNet      & $75.66$          & $74.59$ & $75.44$          &  & $77.07$          & $74.05$ & $73.71$        \\
            & EfficientNetB0 & $75.62$          &         &                  &  & $\mathbf{92.94}$ &         &                \\
            & EfficientNetB1 & $75.11$          &         &                  &  & $91.21$          &         &                \\
            & ResNet34       & $\mathbf{76.09}$ &         &                  &  & $85.41$          &         &                
\end{tblr}

\end{table*}

\renewcommand\arraystretch{1.2}

\begin{table*}[htbp]
\centering
\caption{Imperceptibility in the all-to-all setting. $\uparrow$ means larger values are preferred, and vice versa. For each dataset, the row settings are the same as \cref{acc for all2all}.}
\label{imp for all2all}
\resizebox{1\columnwidth}{!}{
\LARGE
\begin{tblr}{
  row{odd} = {c},
  row{2} = {c},
  row{4} = {c},
  row{8} = {c},
  row{12} = {c},
  cell{1}{1} = {r=2}{},
  cell{1}{2} = {c=3}{},
  cell{1}{6} = {c=3}{},
  cell{1}{10} = {c=3}{},
  cell{1}{14} = {c=3}{},
  cell{1}{18} = {c=3}{},
  cell{3}{1} = {r=4}{},
  cell{3}{3} = {r=4}{},
  cell{3}{4} = {r=4}{},
  cell{3}{7} = {r=4}{},
  cell{3}{8} = {r=4}{},
  cell{3}{11} = {r=4}{},
  cell{3}{12} = {r=4}{},
  cell{3}{15} = {r=4}{},
  cell{3}{16} = {r=4}{},
  cell{3}{19} = {r=4}{},
  cell{3}{20} = {r=4}{},
  cell{6}{2} = {c},
  cell{6}{6} = {c},
  cell{6}{10} = {c},
  cell{6}{14} = {c},
  cell{6}{18} = {c},
  cell{7}{1} = {r=4}{},
  cell{7}{3} = {r=4}{},
  cell{7}{4} = {r=4}{},
  cell{7}{7} = {r=4}{},
  cell{7}{8} = {r=4}{},
  cell{7}{11} = {r=4}{},
  cell{7}{12} = {r=4}{},
  cell{7}{15} = {r=4}{},
  cell{7}{16} = {r=4}{},
  cell{7}{19} = {r=4}{},
  cell{7}{20} = {r=4}{},
  cell{10}{2} = {c},
  cell{10}{6} = {c},
  cell{10}{10} = {c},
  cell{10}{14} = {c},
  cell{10}{18} = {c},
  cell{11}{1} = {r=4}{},
  cell{11}{3} = {r=4}{},
  cell{11}{4} = {r=4}{},
  cell{11}{7} = {r=4}{},
  cell{11}{8} = {r=4}{},
  cell{11}{11} = {r=4}{},
  cell{11}{12} = {r=4}{},
  cell{11}{15} = {r=4}{},
  cell{11}{16} = {r=4}{},
  cell{11}{19} = {r=4}{},
  cell{11}{20} = {r=4}{},
  cell{14}{2} = {c},
  cell{14}{6} = {c},
  cell{14}{10} = {c},
  cell{14}{14} = {c},
  cell{14}{18} = {c},
  hline{1,15} = {-}{0.08em},
  hline{2} = {2-4,6-8,10-12,14-16,18-20}{},
  hline{3,7,11} = {-}{},
}
          & CIEDE2000$\downarrow$ &          &                &  & PSNR(dB)$\uparrow$ &         &                &  & SSIM$(\times 100)\uparrow$ &         &                  &  & $\ell_2\downarrow$ &        &                &  & $\ell_\infty\downarrow$ &        &                 \\
          & Impart                & WaNet    & $\text{Bpp}^*$ &  & Impart             & WaNet   & $\text{Bpp}^*$ &  & Impart                     & WaNet   & $\text{Bpp}^*$   &  & Impart             & WaNet  & $\text{Bpp}^*$ &  & Impart                  & WaNet  & $\text{Bpp}^*$  \\
CIFAR-10  & $\mathbf{21.48}$      & $121.73$ & $118.14$       &  & $42.60$            & $28.39$ & $33.37$        &  & $\mathbf{99.43}$           & $93.81$ & $98.86$          &  & $\mathbf{0.43}$    & $2.23$ & $1.24$         &  & $\mathbf{0.06}$         & $0.27$ & $0.11$          \\
          & $23.12$               &          &                &  & $42.20$            &         &                &  & $99.21$                    &         &                  &  & $0.50$             &        &                &  & $0.07$                  &        &                 \\
          & $25.33$               &          &                &  & $\mathbf{42.81}$   &         &                &  & $99.24$                    &         &                  &  & $0.49$             &        &                &  & $\mathbf{0.06}$         &        &                 \\
          & $22.10$               &          &                &  & $42.33$            &         &                &  & $99.40$                    &         &                  &  & $0.45$             &        &                &  & $\mathbf{0.06}$         &        &                 \\
GTSRB     & $28.57$               & $106.05$ & $96.31$        &  & $\mathbf{47.47}$   & $31.52$ & $39.34$        &  & $98.54$                    & $94.96$ & $\mathbf{98.81}$ &  & $\mathbf{0.52}$    & $1.85$ & $0.60$         &  & $0.09$                  & $0.23$ & $\mathbf{0.03}$ \\
          & $27.97$               &          &                &  & $41.33$            &         &                &  & $98.43$                    &         &                  &  & $0.57$             &        &                &  & $0.08$                  &        &                 \\
          & $\mathbf{26.63}$      &          &                &  & $41.10$            &         &                &  & $98.61$                    &         &                  &  & $0.54$             &        &                &  & $0.08$                  &        &                 \\
          & $29.88$               &          &                &  & $43.78$            &         &                &  & $98.32$                    &         &                  &  & $0.60$             &        &                &  & $0.09$                  &        &                 \\
CIFAR-100 & $\mathbf{22.17}$      & $116.59$ & $106.13$       &  & $\mathbf{42.15}$   & $28.92$ & $34.24$        &  & $\mathbf{99.24}$           & $94.83$ & $98.93$          &  & $\mathbf{0.45}$    & $2.13$ & $1.13$         &  & $\mathbf{0.06}$         & $0.26$ & $0.10$          \\
          & $32.72$               &          &                &  & $39.88$            &         &                &  & $98.81$                    &         &                  &  & $0.61$             &        &                &  & $0.07$                  &        &                 \\
          & $35.08$               &          &                &  & $39.31$            &         &                &  & $98.68$                    &         &                  &  & $0.64$             &        &                &  & $0.07$                  &        &                 \\
          & $25.11$               &          &                &  & $41.04$            &         &                &  & $99.11$                    &         &                  &  & $0.52$             &        &                &  & $\mathbf{0.06}$         &        &                 
\end{tblr}
}

\end{table*}

\noindent\textbf{Attack Setup.} 
We set the poisoned ratio $\rho=$10\% for all experiments if not specified otherwise. Besides, we train the networks using SGD optimizer with weight decay $5\times 10^{-4}$. The learning rate is using cosine decay with a warm-up. The upper bound of the learning rate is set at $0.01$. We set $\gamma=10$ to \cref{loss}.

\noindent\textbf{Evaluation Metrics.}
To evaluate the attack effectiveness, we use the attack success ratio (ASR) and benign accuracy(BA) designed in the previous works~\citep{li2021invisible, nguyen2021wanet, wang2022bppattack}. Besides, we use the accuracy of the benign model(ACC) as a reference. In detail, the ACC evaluates the accuracy of the benign model on benign data while the BA evaluates the accuracy of the poisoned model on benign data. The ACC is shown in \cref{dataset}.

In order to comprehensively evaluate the quality of the poisoned images, We select five image quality metrics CIEDE2000~\citep{luo2001development}, SSIM~\citep{wang2004image}, PSNR~\citep{huynh2008scope}, $\ell_2$ and $\ell_\infty$. Both SSIM and CIEDE2000 are designed to better align with human visual perception, where CIEDE2000 is prone to color discrepancy while SSIM extracts structural information from the viewing field.

\begin{table*}[htbp]
\centering
\caption{Attack effectiveness in the all-to-one setting. For each dataset, four rows represent the SM as GoogleNet, EfficientNetB0, EfficientNetB1, and ResNet34, respectively. $*$ means the poisoned ratio is 20\%, and the others is 10\%.
}
\label{acc for all2one}
\begin{tblr}{
  row{odd} = {c},
  row{2} = {c},
  row{4} = {c},
  row{8} = {c},
  row{12} = {c},
  cell{1}{2} = {r=2}{},
  cell{1}{3} = {c=3}{},
  cell{1}{7} = {c=3}{},
  cell{3}{1} = {r=4}{},
  cell{3}{4} = {r=4}{},
  cell{3}{5} = {r=4}{},
  cell{3}{8} = {r=4}{},
  cell{3}{9} = {r=4}{},
  cell{6}{2} = {c},
  cell{6}{3} = {c},
  cell{6}{7} = {c},
  cell{7}{1} = {r=4}{},
  cell{7}{4} = {r=4}{},
  cell{7}{5} = {r=4}{},
  cell{7}{8} = {r=4}{},
  cell{7}{9} = {r=4}{},
  cell{10}{2} = {c},
  cell{10}{3} = {c},
  cell{10}{7} = {c},
  cell{11}{1} = {r=4}{},
  cell{11}{4} = {r=4}{},
  cell{11}{5} = {r=4}{},
  cell{11}{8} = {r=4}{},
  cell{11}{9} = {r=4}{},
  cell{14}{2} = {c},
  cell{14}{3} = {c},
  cell{14}{7} = {c},
  hline{1,15} = {-}{0.08em},
  hline{2} = {3-5,7-9}{},
  hline{3,7,11} = {-}{},
}
            & SM             & BA(\%)           &         &                  &  & ASR(\%)        &         &                  \\
            &                & Impart           & WaNet   & $\text{Bpp}^*$   &  & Impart         & WaNet   & $\text{Bpp}^*$   \\
CIFAR-$10$  & GoogleNet      & $94.11$          & $94.15$ & $\mathbf{94.54}$ &  & $98.97$        & $99.55$ & $\mathbf{99.91}$ \\
            & EfficientNetB0 & $94.37$          &         &                  &  & $98.07$        &         &                  \\
            & EfficientNetB1 & $94.12$          &         &                  &  & $98.01$        &         &                  \\
            & ResNet34       & $94.22$          &         &                  &  & $99.03$        &         &                  \\
GTSRB       & GoogleNet      & $98.77$          & $98.97$ & $\mathbf{99.25}$ &  & $99.70$        & $98.78$ & $99.96$          \\
            & EfficientNetB0 & $98.79$          &         &                  &  & $\mathbf{100}$ &         &                  \\
            & EfficientNetB1 & $98.73$          &         &                  &  & $\mathbf{100}$ &         &                  \\
            & ResNet34       & $98.08$          &         &                  &  & $99.89$        &         &                  \\
CIFAR-$100$ & GoogleNet      & $75.66$          & $75.40$ & $\mathbf{76.55}$ &  & $99.61$        & $99.56$ & $\mathbf{100}$   \\
            & EfficientNetB0 & $\mathbf{76.55}$ &         &                  &  & $99.58$        &         &                  \\
            & EfficientNetB1 & $76.51$          &         &                  &  & $99.43$        &         &                  \\
            & ResNet34       & $76.07$          &         &                  &  & $99.81$        &         &                  
\end{tblr}

\end{table*}
\begin{table*}[htbp]
\centering
\caption{Imperceptibility in the all-to-one setting. $\uparrow$ means larger values are preferred, and vice versa. For each dataset, the row settings are the same as \cref{acc for all2one}.}
\label{imp for all2all}
\resizebox{1\columnwidth}{!}{
\LARGE

\begin{tblr}{
  cells = {c},
  cell{1}{1} = {r=2}{},
  cell{1}{2} = {c=3}{},
  cell{1}{6} = {c=3}{},
  cell{1}{10} = {c=3}{},
  cell{1}{14} = {c=3}{},
  cell{1}{18} = {c=3}{},
  cell{3}{1} = {r=4}{},
  cell{3}{3} = {r=4}{},
  cell{3}{4} = {r=4}{},
  cell{3}{7} = {r=4}{},
  cell{3}{8} = {r=4}{},
  cell{3}{11} = {r=4}{},
  cell{3}{12} = {r=4}{},
  cell{3}{15} = {r=4}{},
  cell{3}{16} = {r=4}{},
  cell{3}{19} = {r=4}{},
  cell{3}{20} = {r=4}{},
  cell{6}{6} = {},
  cell{6}{10} = {},
  cell{6}{14} = {},
  cell{6}{18} = {},
  cell{7}{1} = {r=4}{},
  cell{7}{3} = {r=4}{},
  cell{7}{4} = {r=4}{},
  cell{7}{7} = {r=4}{},
  cell{7}{8} = {r=4}{},
  cell{7}{11} = {r=4}{},
  cell{7}{12} = {r=4}{},
  cell{7}{15} = {r=4}{},
  cell{7}{16} = {r=4}{},
  cell{7}{19} = {r=4}{},
  cell{7}{20} = {r=4}{},
  cell{10}{2} = {},
  cell{10}{6} = {},
  cell{10}{10} = {},
  cell{10}{14} = {},
  cell{10}{18} = {},
  cell{11}{1} = {r=4}{},
  cell{11}{3} = {r=4}{},
  cell{11}{4} = {r=4}{},
  cell{11}{7} = {r=4}{},
  cell{11}{8} = {r=4}{},
  cell{11}{11} = {r=4}{},
  cell{11}{12} = {r=4}{},
  cell{11}{15} = {r=4}{},
  cell{11}{16} = {r=4}{},
  cell{11}{19} = {r=4}{},
  cell{11}{20} = {r=4}{},
  cell{14}{2} = {},
  cell{14}{6} = {},
  cell{14}{10} = {},
  cell{14}{14} = {},
  cell{14}{18} = {},
  hline{1,3,7,11,15} = {-}{},
  hline{2} = {2-4,6-8,10-12,14-16,18-20}{},
}
          & CIEDE2000$\downarrow$ &          &                &  & PSNR(dB)$\uparrow$ &         &                &  & SSIM$(\times 100)\uparrow$ &         &                  &  & $\ell_2\downarrow$ &        &                &  & $\ell_\infty\downarrow$ &        &                 \\
          & Impart                & WaNet    & $\text{Bpp}^*$ &  & Impart             & WaNet   & $\text{Bpp}^*$ &  & Impart                     & WaNet   & $\text{Bpp}^*$   &  & Impart             & WaNet  & $\text{Bpp}^*$ &  & Impart                  & WaNet  & $\text{Bpp}^*$  \\
CIFAR-10  & $20.36$               & $121.73$ & $118.14$       &  & $53.11$            & $28.39$ & $33.37$        &  & $\mathbf{99.45}$           & $93.81$ & $98.86$          &  & $\mathbf{0.40}$    & $2.23$ & $1.24$         &  & $\mathbf{0.05}$         & $0.27$ & $0.11$          \\
          & $21.36$               &          &                &  & $52.66$            &         &                &  & $99.26$                    &         &                  &  & $0.47$             &        &                &  & $0.06$                  &        &                 \\
          & $24.02$               &          &                &  & $52.99$            &         &                &  & $99.27$                    &         &                  &  & $0.46$             &        &                &  & $\mathbf{0.05}$         &        &                 \\
          & $\mathbf{20.28}$      &          &                &  & $\mathbf{53.14}$   &         &                &  & $99.44$                    &         &                  &  & $0.41$             &        &                &  & $\mathbf{0.05}$         &        &                 \\
GTSRB     & $\mathbf{29.05}$               & $106.05$ & $96.31$        &  & $\mathbf{48.24}$   & $31.52$ & $39.34$        &  & $98.51$                    & $94.96$ & $\mathbf{98.81}$ &  & $\mathbf{0.52}$    & $1.85$ & $0.60$         &  & $0.08$                  & $0.23$ & $\mathbf{0.03}$ \\
          & $29.12$               &          &                &  & $41.92 $           &         &                &  & $98.31$                    &         &                  &  & $0.60$             &        &                &  & $0.09$                  &        &                 \\
          & $29.36$               &          &                &  & $41.22$            &         &                &  & $98.34$                    &         &                  &  & $0.59$             &        &                &  & $0.08$                  &        &                 \\
          & $30.18$      &          &                &  & $44.07$            &         &                &  & $98.32$                    &         &                  &  & $0.61$             &        &                &  & $0.09$                  &        &                 \\
CIFAR-100 & $\mathbf{23.69}$      & $116.59$ & $106.13$       &  & $\mathbf{42.61}$   & $28.92$ & $34.24$        &  & $\mathbf{99.17}$           & $94.83$ & $98.93$          &  & $\mathbf{0.48}$    & $2.13$ & $1.13$         &  & $\mathbf{0.06}$         & $0.26$ & $0.10$          \\
          & $34.56$               &          &                &  & $40.56$            &         &                &  & $98.85$                    &         &                  &  & $0.61$             &        &                &  & $0.09$                  &        &                 \\
          & $37.88$               &          &                &  & $39.88$            &         &                &  & $98.67$                    &         &                  &  & $0.67$             &        &                &  & $0.09$                  &        &                 \\
          & $26.76$               &          &                &  & $41.51$            &         &                &  & $99.02$                    &         &                  &  & $0.54$             &        &                &  & $\mathbf{0.06}$         &        &                 
\end{tblr}
}
\end{table*}
\subsection{Attack Experiments}
\begin{figure*}[htbp]
\centering
\begin{minipage}[t]{0.3\textwidth}
\centering
\includegraphics[width=5cm]{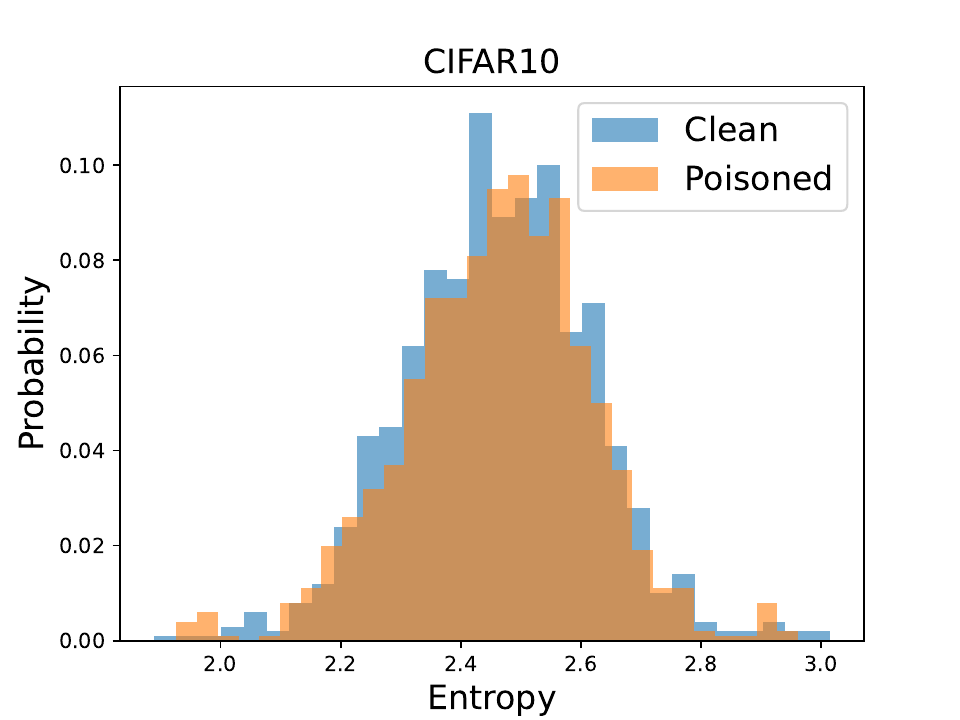}
\end{minipage}
\begin{minipage}[t]{0.3\textwidth}
\centering
\includegraphics[width=5cm]{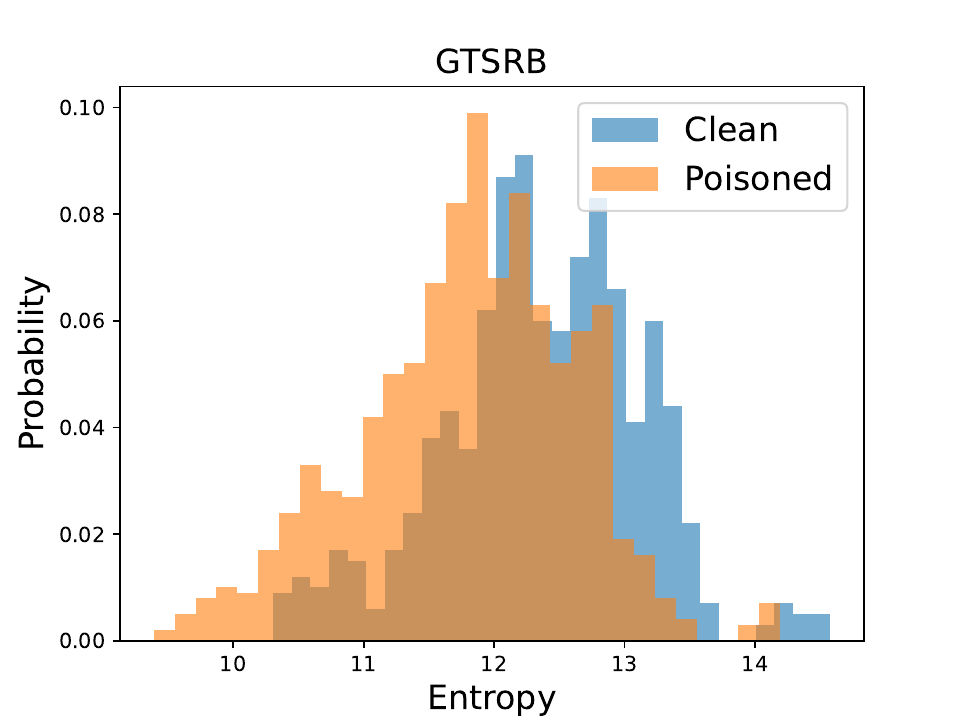}
\end{minipage}
\begin{minipage}[t]{0.3\textwidth}
\centering
\includegraphics[width=5cm]{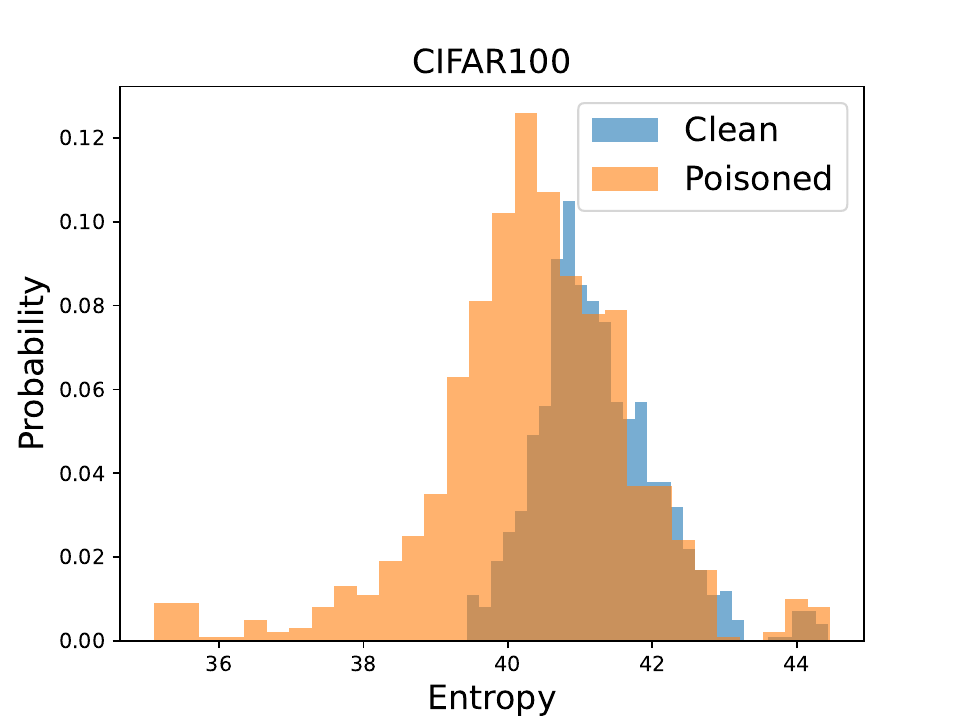}
\end{minipage}
\caption{Resilience to STRIP. Entropy distributions on CIFAR-10, GTSRB and CIFAR-100. }
\label{strip_all2all}
\end{figure*}
For the all-to-all setting, \cref{acc for all2all} shows that Impart outperforms the baselines in CIFAR-10 and CIFAR-100 on ASR without scarifying BA. Meanwhile, it achieves comparable results with baselines in GTSRB. Especially, for the GTSRB dataset, as described in the \cref{sec:intro}, just because the ACC of the GTSRB dataset in PreActResNet18 is about 99\%, hence all the previous works also can achieve about 99\% ASR in this dataset. However, our method achieves 92.94\% ASR on CIFAR-100 when using EfficientNetB0 as the surrogate model which is much higher than the baseline. This is because the generated poisoned images in Impart are associated with the target label before training, which makes it much easier to learn the mapping from triggers to the target label than baselines which only have random perturbations. In the CIFAR-100 dataset when using GoogleNet as the surrogate model, the ASR is 77.07\%. It is because the fitting ability of the GoogleNet is worst than the EfficientNetB0 in the CIFAR-100 dataset. The model EfficientNetB0 can generate more precise perturbation which is related to the target label than the model GoogleNet. Finally, the victim model can easier learn the poisoned examples generated by EfficientNetB0 more than GoogleNet. To sum up, our method is more applicable to real-world datasets rather than the datasets like GTSRB which can achieve about 99\% ACC.
Besides, \cref{imp for all2all} shows Impart achieves remarkable superior image quality overall. In detail, when we compare SSIM only, we observe that the SSIM of Bpp is comparable with that of Impart. This indeed verifies Bpp retains the structure information while losing the color information because of using fewer bits to describe a pixel. Meanwhile, Impart holds the structure information and color information simultaneously, and thus, on CIEDE2000 which is a metric considering color information, Impart is far better than Bpp. 

In addition, for the all-to-one setting as shown in \cref{acc for all2one}, all the attack methods achieve high attack performance, meanwhile Impart also achieves remarkable overall advantages on image quality compared with baseline methods as shown in \cref{imp for all2all}.  It is noted that since the poison images of baselines remain the same for both all-to-all and all-to-one settings, their image qualities do not change.  However, the poison images of Impart rely on the label, and so the qualities under different settings are different.  \cref{acc for all2one} shows that ASR of Impart is comparable but slightly worse than baselines, especially Bpp whose poison ratio is 20\%. We recognize that as introduced in \cref{sec:intro}, the task of the all-to-one setting is simpler than that of the all-to-all because it only addresses one mapping from triggers to the target label. All methods including Impart can handle it well.  However, since Impart has more strict requirements on image imperceptibility, it scarifies the effectiveness slightly.  Thus, Impart can still achieve a comparable performance even if it is designed for the all-to-all setting.
\subsection{Defense Experiments}

\begin{figure}
\centering
\includegraphics[width=6.5cm]{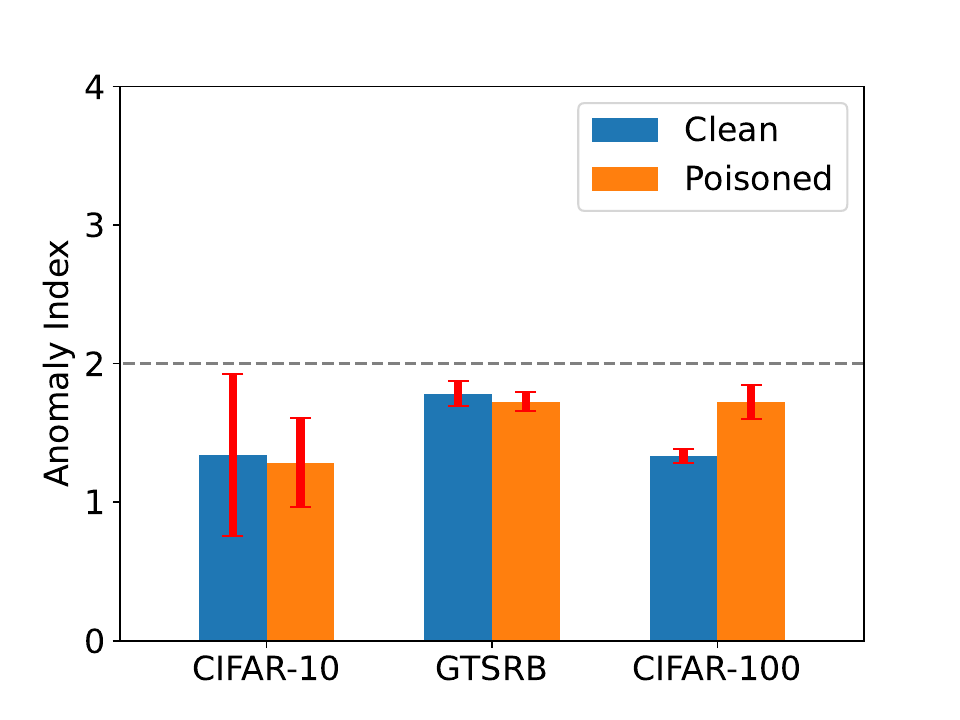}
\caption{Effectiveness of our method under the Neural Cleanse defense on CIFAR-10, GTSRB, and CIFAR-100 datasets.}
\label{nc_all2all}
\end{figure}

In this section, we investigate whether our method can bypass existing state-of-the-art defense methods: Neural Cleanse~\citep{wang2019neural}, STRIP~\citep{gao2019strip}, Neural Attention Distillation~\citep{li2021neural}, Spectral Signatures~\citep{BrandonTran2018SpectralSI} and GradCam~\citep{selvaraju2017grad}. We select these defense methods because they are widely used in previous works~\citep{doan2021lira, doan2021backdoor}. Besides, they evaluate the backdoor attack method from different aspects.
We conduct the defense methods in the all-to-all setting with GoogleNet as the surrogate model. 
Additionally, all of the defense methods are conducted in the default hyperparameters declared in their original papers.

\noindent\textbf{Neural Cleanse.}
Neural Cleanse (NC)~\citep{wang2019neural} is a typical trigger synthesis-based empirical defense method, which reconstructs the triggers through reverse engineering. 
It firstly treats each class label as a potential target label, then reconstructs a ``minima'' trigger that can produce model misclassification by a designed optimization.  Finally, it runs an outlier detection to find the anomaly based on the assumption that the infected label requires much smaller modifications to cause misclassification than other uninfected labels. 

\cref{nc_all2all} shows the resistance results to NC. It shows that the anomaly index of our method computed by NC is lower than the threshold value 2. It states Impart passes the defense on all datasets. While previous works WaNet and Bpp also pass against NC defense, they implement negative (noise) sample (i.e., 20\% negative sample in WaNet and Bpp) as auxiliary which significantly reduces the quality of the dataset. As shown in \cref{imp for all2all}, PSNR of WaNet is 29.61dB and Bpp is 36.65dB on average. On the contrary,  Impart is robust against NC just because the generated perturbation is minor and uniform for all classes.

\begin{figure*}[t]
\centering
\begin{minipage}[t]{0.3\textwidth}
\centering
\includegraphics[width=5cm]{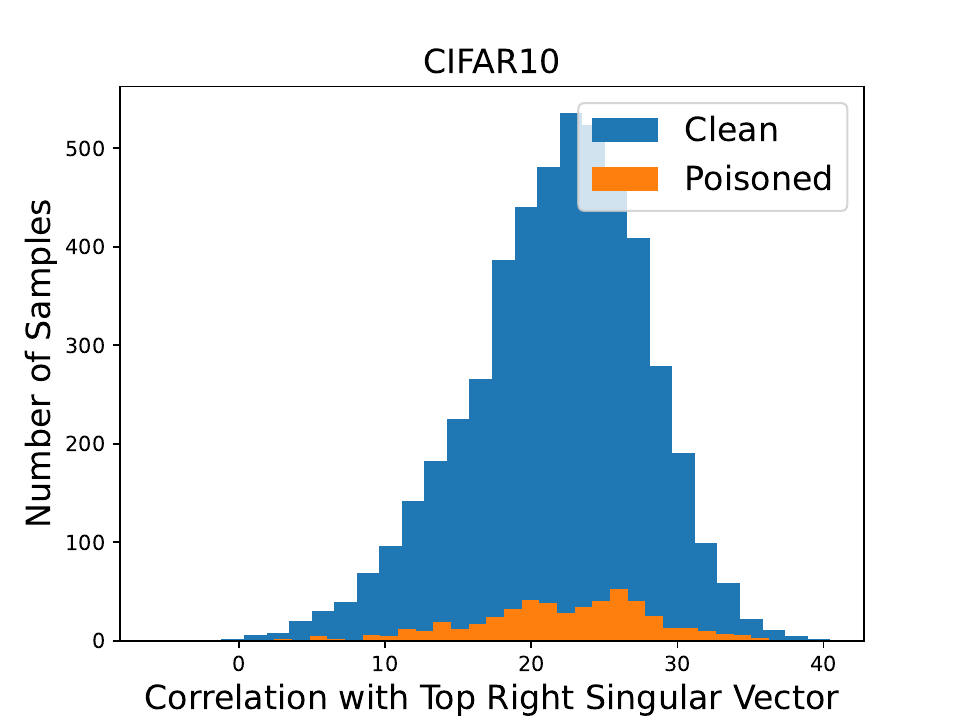}
\end{minipage}
\begin{minipage}[t]{0.3\textwidth}
\centering
\includegraphics[width=5cm]{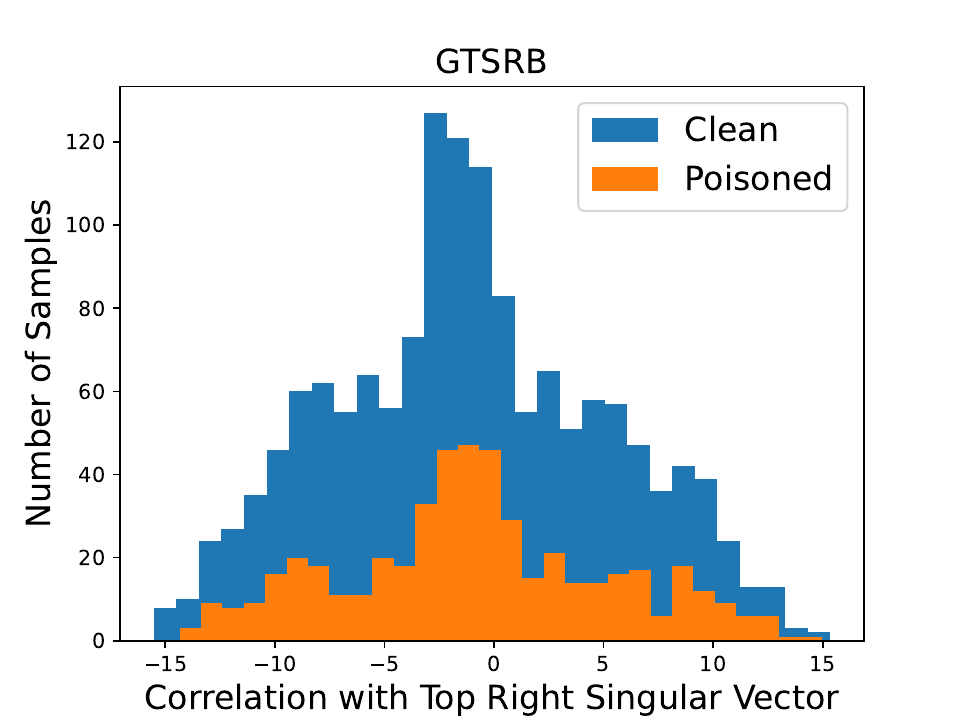}
\end{minipage}
\begin{minipage}[t]{0.3\textwidth}
\centering
\includegraphics[width=5cm]{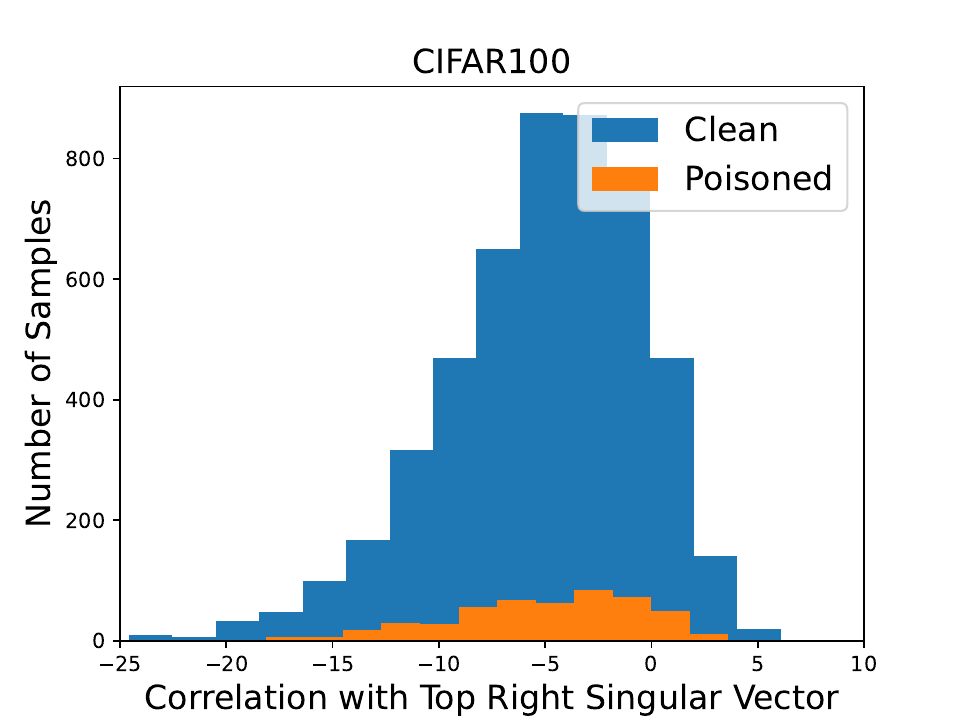}
\end{minipage}
\caption{Resilience to Spectral Signatures.}
\label{ss_all2all}
\end{figure*}

\noindent\textbf{STRIP.}
We then select a representative sample filtering-based empirical defense STRIP~\citep{gao2019strip}. It determines the presence of the backdoor by calculating the entropy predicted by the model before and after perturbing the inputs intentionally. The entropy of the input below 0.2 states this image is a poisoned sample. We apply this method and calculate the minimal entropy with respect to three datasets individually, which are 1.88, 9.39, and 35.09, respectively. As all of them are far larger than 0.2,  this shows Impart is resistant to STRIP. Besides,  \cref{strip_all2all} shows that the entropy range is similar between the benign model and the poisoned model, and the entropy range of the poisoned model includes the entropy range of the benign model. For the result of the GTSRB dataset, there are some separate in our view, but for a defender who doesn't know the existence of poison data, he can't distinguish there are two different distributions.
These results indicate our method can successfully bypass the STRIP defense method.

\renewcommand\arraystretch{1}
\begin{table}[htbp]
\centering
\caption{\centering Effectiveness of our method under NAD on three datasets.}
\label{nad_all2all}
\begin{tabular}{ccccc} 
\toprule
\multirow{2}{*}{Datasets} & \multicolumn{2}{c}{No Defense} & \multicolumn{2}{c}{NAD}  \\ 
                          & BA(\%) & ASR(\%)                       & BA(\%) & ASR(\%)                 \\ 
\hline
CIFAR-10                  & $94.24$  & $96.35$                      & $35.20$  & $7.30$                   \\
GTSRB                     & $98.03$  & $98.90$                      & $18.30$  & $8.23$                   \\
CIFAR-100                 & $75.66$  & $77.07$                      & $4.86$  & $0.75$                   \\
\bottomrule
\end{tabular}

\end{table}

\begin{table}
\centering
\begin{minipage}[t]{0.4\textwidth}
\centering
\includegraphics[width=6.5cm]{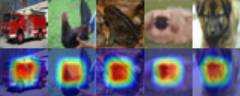}
\textbf{Clean}
\end{minipage}

\begin{minipage}[t]{0.4\textwidth}
\centering
\includegraphics[width=6.5cm]{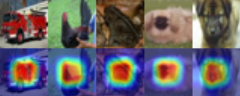}
\textbf{Poisoned}
\end{minipage}
\caption{Resilience to GradCam. {Comparisons of GradCam visualization maps between clean data ($1^{st}$ row) and poisoned data ($2^{nd}$ row) on a poisoned model.}}
\label{gradcam_all2all}
\end{table}
\noindent\textbf{Neural Attention Distillation.}
Neural Attention Distillation(NAD)~\citep{li2021neural} is a state-of-the-art model reconstruction-based empirical defense. 
NAD erases the backdoor through a twice-finetuning process. 
It firstly gets a teacher model by finetuning the victim model on a small clean subset, then utilizing the teacher model to guide the second finetuning aiming to the selected layer attention distribution of the student model align to the teacher model. 
As shown in \cref{nad_all2all}, after the distillation, BA decreases dramatically. This means that NAD is ineffective against our method Impart. We speculate it may be because the generated poisoning images of our method is aligned with the target class in the image feature, the NAD can't distinguish the poison data and clean data.

\noindent\textbf{Spectral Signatures.}
Spectral Signatures~\citep{BrandonTran2018SpectralSI} is a representative defensive approach that inspects the latent space of the model. It first finds the top-right singular vector of the covariance matrix of the latent vectors using a small subset of clean samples. Then it calculates the correlation of each sample to this singular vector. Those with the outlier scores are identified as backdoor samples. \cref{ss_all2all} shows our method can bypass it as the distributions are similar. This is because we use a clean surrogate model to generate poison examples that are related to the target label, and so the latent space representation is also imparted to the victim model by the poison examples.

\noindent\textbf{GradCam Visualization.}
GradCam~\citep{selvaraju2017grad} is a typical network visualization method. 
It reveals the model behaviors by a heat map of the input image. The heat map denotes the attention when the model recognizes the input image. 
Following the previous work~\citep{nguyen2021wanet}, we compare the attention discrepancy of the poisoned model between benign data and poisoned data. 
The inputs use the ground truth label for benign data while the target label is for poisoned data. We use the existing package pytorch-grad-cam~\citep{jacobgilpytorchcam} to evaluate our method.  
\cref{gradcam_all2all}  shows that the heat maps produced by GradCam are nearly identical between benign data and poisoned data. 

\subsection{Ablation Studies}



\begin{table*}[htbp]
\centering
\caption{\centering The influence of different $\gamma$. $\uparrow$ indicates larger values are preferred, and vise versa.}
\label{mulgamma}
\begin{tabular}{cccccc} 
\toprule
\diagbox{Metrics}{$\gamma$} & $100$    & $50$    & $30$    & $10$    & $0$      \\ 
\hline
BA(\%)                       & $98.96$  & $96.96$ & $95.9$  & $98.03$ & $98.75$  \\
ASR(\%)                      & $55.76$  & $85.35$ & $95.92$ & $98.90$ & $98.51$  \\
\hline
CIEDE2000$\downarrow$                    & $4.31$   & $11.52$    & $18.47$    & $28.56$    & $29.85$  \\
PSNR(dB)$\uparrow$                      & $125.59$ & $84.32$ & $64.55$ & $47.47$ & $40.23$     \\
SSIM$(\times100)\uparrow$                          & $99.76$  & $99.44$    & $99.16$    & $98.54$    & $98.10$    \\
$\ell_2\downarrow$                      & $0.06$   & $0.16$     & $0.28$     & $0.52$     & $0.69$     \\
$\ell_\infty\downarrow$                & $0.01$   & $0.03$     & $0.05$     & $0.09$     & $0.14$     \\
\bottomrule
\end{tabular}

\end{table*}
\label{gamma}
We investigate the influence of hyperparameters $\rho$ and $\gamma$. All experiments are conducted in the all-to-all setting.

%

\begin{figure}
\centering
\begin{minipage}[t]{0.4\textwidth}
\centering
\includegraphics[width=6cm]{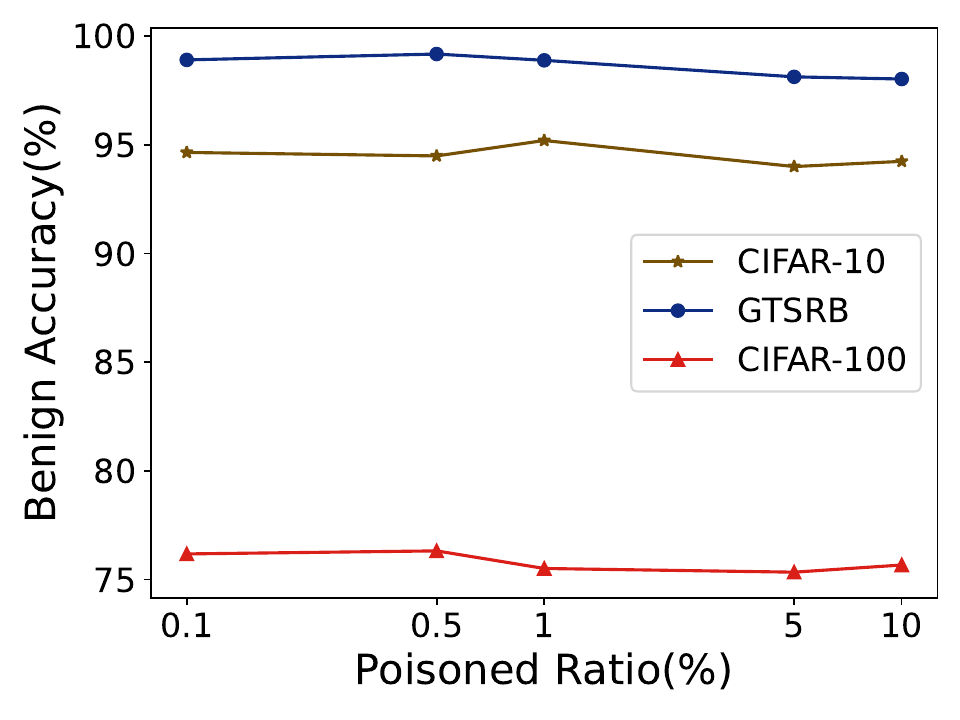}

(a) BA
\end{minipage}
\begin{minipage}[t]{0.4\textwidth}
\centering
\includegraphics[width=6cm]{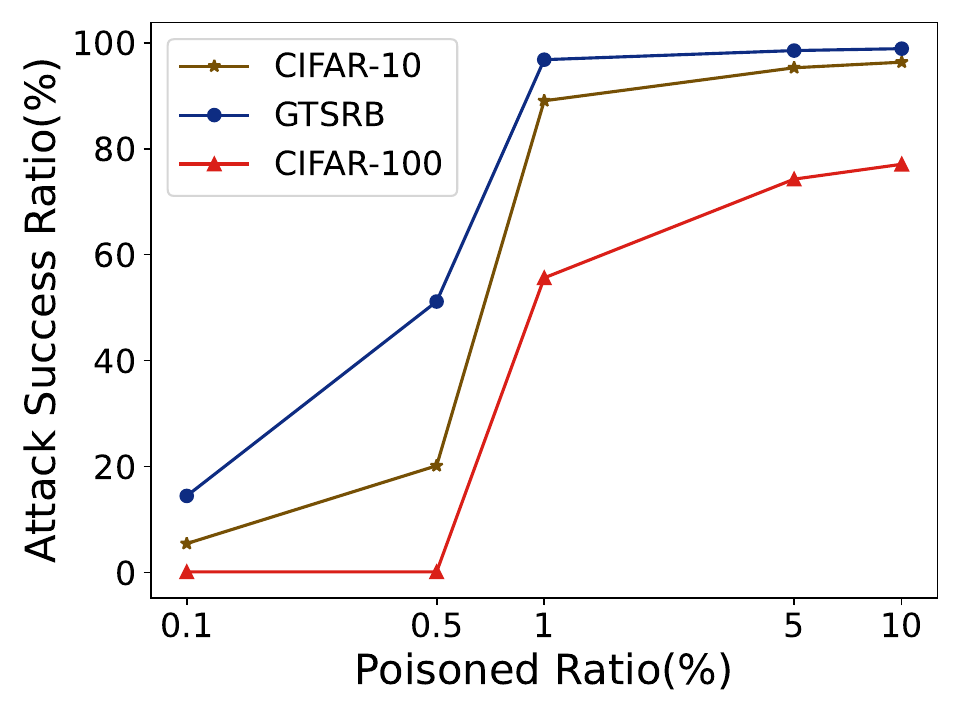}

(b) ASR
\end{minipage}
\caption{The influence of different poisoned ratio. The left (a) is the influence of poisoned ratio on BA. The right (b) is the influence of poisoned ratio on ASR.}
\label{poisoned ratio}
\end{figure}
\noindent\textbf{Poisoned Ratio $\rho$.}
We evaluate the influence of poisoned ratio $\rho$ following the setting of previous work~\citep{wu2022backdoorbench} that recorded the BA and ASR with poisoned ratios 10\%, 5\%, 1\%, 0.5\%, 0.1\%  for all three datasets. \cref{poisoned ratio}(a) reveals there has been a steady trend for BA in all datasets. Additionally, \cref{poisoned ratio}(b) shows there has been a sharp drop for ASR in a range from 1\% to 0.1\% but a slight decline in a range from 10\% to 1\%. What is striking in this figure is that our method still achieves high ASR only with a 1\% poisoned ratio. This indeed declared the effectiveness of Impart.



\begin{figure}[htbp]
\centering
\begin{minipage}[t]{0.4\textwidth}
\centering
\includegraphics[width=5.5cm]{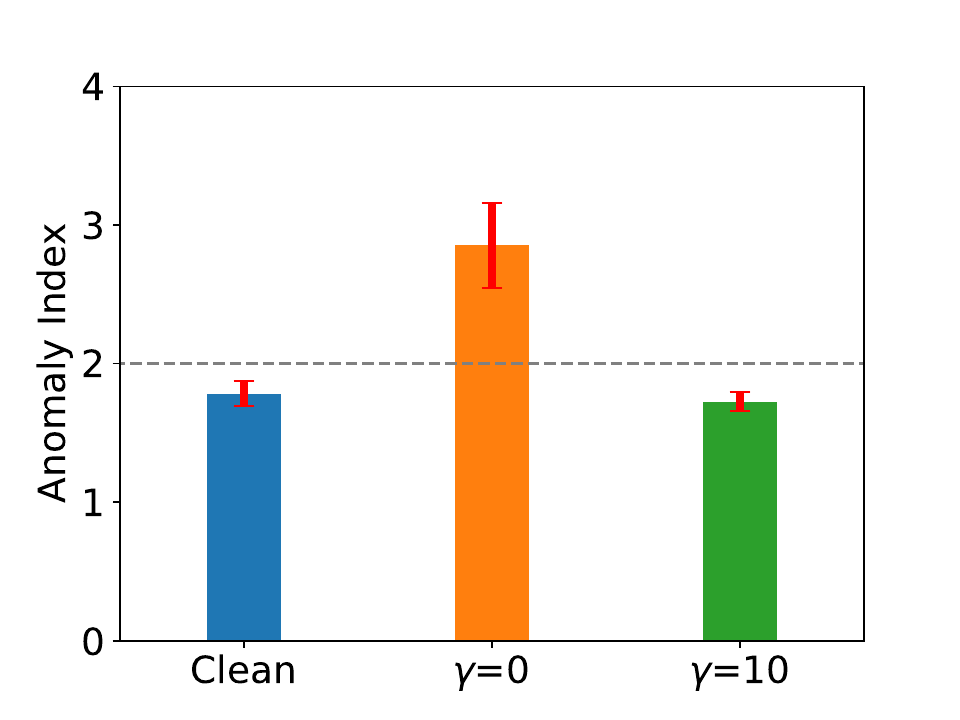}

(a) NC
\end{minipage}
\begin{minipage}[t]{0.4\textwidth}
\centering
\includegraphics[width=2.5cm]{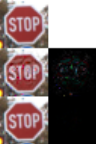}

(b) Worst case
\end{minipage}
\caption{Effectiveness of $\ell_2$ regularization. (a) The effectiveness of $\gamma$ resistance to NC defense. (b) First row:clean image; second row: $\gamma=0$; third row: $\gamma=10$. }

\label{gamma_all2all}
\end{figure}
\noindent\textbf{Hyperparameter $\gamma$.}
To investigate the effectiveness of hyperparameter $\gamma$, we conduct comprehensive experiments with five different $\gamma=0,10,30,50$ and $100$ in the GTSRB dataset.  The iteration number of generating adversarial examples is set to 200. \cref{mulgamma} shows the correlation among $\gamma$, image quality metrics, ASR and BA. It shows that both the image quality and the ASR drop with the increase of $\gamma$. Besides, the BA is relatively steady alongside the increase of $\gamma$. Additionally, we can observe from \cref{mulgamma} that the larger the $\gamma$, the better the image quality. In other words, we can infer that the larger the $\gamma$, the less iteration time to achieve identical image quality.

Furthermore, \cref{gamma_all2all}(a) shows the effectiveness of $\gamma$ resistance to NC. \cref{gamma_all2all} (b) shows the influence of image quality with and without using $\ell_2$ in the worst case. 
We observe a difference between the clean image and the poisoned image when $\gamma=0$ but they are similar when $\gamma=10$. 

\subsection{Discussion}
\noindent\textbf{Compared with White-Box methods in Tiny-Imagenet.} To further demonstrate the attack effectiveness of our method, we compare our method Impart with the White-Box method LIRA~\citep{doan2021lira} and WB~\citep{doan2021backdoor} in the Tiny-Imagenet dataset. Besides, because the generalization error in the Tiny-Imagenet dataset is too large to regard as a benign model(i.e., train accuracy: 99.28\% and test accuracy: 59.12\%), so we decide to list the results here as a case study. As for the victim model(VM), we use ResNet18. For the surrogate model(SM) used to fit the image feature, we select  EfficientNetB0 since it is more effective to fit the image feature shown in the previous experiments.
 The results are shown in \cref{etdataset}, which declares that our method can achieve about 74\% ASR in the all-to-all setting but the white-box method LIRA can only achieve 59\%. This indeed demonstrates the effectiveness of our method.

\begin{table*}[htbp]
\centering
\caption{\centering Attack effectiveness in the Tiny Imagenet dataset.}
\label{etdataset}
\begin{tblr}{
  cells = {c},
  cell{1}{1} = {r=2}{},
  cell{1}{2} = {r=2}{},
  cell{1}{3} = {c=2}{},
  cell{1}{6} = {c=2}{},
  cell{1}{9} = {c=2}{},
  hline{1,5} = {-}{0.08em},
  hline{2} = {3-4,6-7,9-10}{},
  hline{3} = {-}{},
}
Dataset    & Mode       & Impart           &                  &  & LIRA    &                &  & WB      &         \\
           &            & BA               & ASR              &  & BA      & ASR            &  & BA      & ASR     \\
T-Imagenet & All-to-All & $\mathbf{59.28}$ & $\mathbf{74.09}$ &  & $58.00$ & $59.00$        &  & $58.00$ & $58.00$ \\
T-Imagenet & All-to-One & $\mathbf{62.13}$ & $99.21$          &  & $58.00$ & $\mathbf{100}$ &  & $57.00$ & $99.00$ 
\end{tblr}

\end{table*}

\noindent\textbf{Adversarial Attack or Backdoor Attack?} One may argue that it is unclear whether the high ASR is due to the backdoor effect of the trigger or the adversarial perturbation. Therefore, we test the adversarial perturbation's effect in the black-box scenario. Specifically, as the standard adversarial attack process, we first use a surrogate model to generate the poison data, then directly feed the poisoned data to the VM (i.e., trained by the clean dataset) in the inference phase to test the ASR. As for the VM, it is set identically with the \cref{sec:exp setup}. For the surrogate model(SM) used to fit the image feature, we select EfficientNetB0 since it is more effective to fit the image feature shown in the previous experiments. The results are shown in \cref{adv}, which declares that the adversarial perturbation is completely ineffective to attack the VM in the black-box scenario. Therefore, we can conclude that the high ASR is indeed due to the backdoor effect of the trigger.

\begin{table}
\centering
\caption{\centering Attack effectiveness of the adversarial perturbation in the black-box scenario.}
\label{adv}
\begin{tblr}{
  cells = {c},
  cell{1}{1} = {r=2}{},
  cell{1}{2} = {c=2}{},
  cell{1}{5} = {c=2}{},
  hline{1,6} = {-}{0.08em},
  hline{2} = {2-3,5-6}{},
  hline{3} = {-}{},
}
            & All-to-All &         &  & All-to-One &         \\
            & BA(\%)     & ASR(\%) &  & BA(\%)     & ASR(\%) \\
CIFAR-$10$  & $94.80 $   & $14.30$ &  & $94.80$    & $17.36$ \\
GTSRB       & $99.23 $   & $9.56$  &  & $99.23$    & $6.46$  \\
CIFAR-$100$ & $76.18$    & $2.86$  &  & $76.18$    & $1.50$  
\end{tblr}
\end{table}

\noindent\textbf{The Selection Strategy of Surrogate Model.} In the paper, we use the existing model as the surrogate model to test the effectiveness of the Impart. Our method requires generating the perturbation related to the target label, thus in order to achieve a high attack success rate 1) the model needs to fit as correctly as possible about the target label, or 2) the model needs to have a similar understanding with the victim model. Therefore, it is better to select the model either 1) the model which has higher accuracy in large-scale and realistic datasets (e.t. ImageNet) or 2) the model which has a more similar structure to the victim model.

\section{Conclusion}
In this study, we propose a novel backdoor attack framework, Impart, that can simultaneously achieve strong attack ability and high imperceptibility without access to the victim model. Different from previous works which try to find an imperceptible pattern and add it to the source image as the poisoned image, we propose to generate perturbation that aligned with the target label in the image feature by a surrogate model. In this way, the generated poisoned images are attached with knowledge about the target class, which significantly enhances the attack capability. We evaluate our method on three benchmark datasets and five typical defense mechanisms. Experiments show that Impart can achieve state-of-the-art attack success rates in the all-to-all setting, meanwhile, it maintains high visual quality. 

\section*{Acknowledgements}
This work is supported by the National Natural Science Foundation of China (62202329).


\bibliographystyle{cas-model2-names}

\bibliography{cas-refs}

\bio{}
\endbio

\endbio

\end{document}